\documentclass[11pt]{article} 

\usepackage[utf8]{inputenc} 
\usepackage{cite}
\usepackage{bbold}
\usepackage{bbm}

\usepackage{geometry} 
\usepackage{graphicx} 
\usepackage{color}
\usepackage{soul}

\usepackage{booktabs} 
\usepackage{array} 
\usepackage{paralist} 
\usepackage{verbatim} 
\usepackage{subfig} 

\usepackage{fancyhdr} 
\usepackage{sectsty}
\allsectionsfont{\sffamily\mdseries\upshape} 

\usepackage[nottoc,notlof,notlot]{tocbibind} 
\usepackage[titles,subfigure]{tocloft} 

\newcommand{\beq}{\begin{equation}}
\newcommand{\eeq}{\end{equation}}

\newcommand{\ie}{{\em i.e., }}
\newcommand{\red}{\color{black}}

\newcommand{\bds}{\begin{displaystyle}}
\newcommand{\eds}{\end{displaystyle}}

\title{A conditional likelihood is required to estimate the selection coefficient in ancient DNA}
\author{Angelo Valleriani\footnote{angelo.valleriani@mpikg.mpg.de}\\
Max Planck Institute of Colloids and Interfaces\\
Department of Theory and Bio-Systems\\
D-14424 Potsdam, Germany}
\date{} 

\begin{document}
\maketitle
\begin{abstract}
Time-series of allele frequencies are a useful and unique set of data to determine the strength of natural selection on the background of genetic drift.  Technically, the selection coefficient is estimated by means of a likelihood function built under the hypothesis that the available trajectory spans a sufficiently large portion of the fitness landscape. Especially for ancient DNA, however, often only one single such trajectories is available and the coverage of the fitness landscape is very limited. In fact, one single trajectory is more representative of a process conditioned both in the initial and in the final condition than of a process free to {\red visit the available fitness landscape}. Based on {\red two models} of population genetics, here we show how to build a likelihood function for the selection coefficient that takes the statistical peculiarity of single trajectories into account. We show that this conditional likelihood delivers a precise estimate of the selection coefficient also when allele frequencies are close to fixation whereas the unconditioned likelihood fails. Finally, we discuss the fact that the traditional, unconditioned likelihood always delivers an answer, which is often unfalsifiable and appears reasonable also when it is {\red not correct}.
\end{abstract} 
\thispagestyle{empty}

\section*{Introduction}
Past records of the frequency of a character, \ie an allele or a phenotype, until present observational time are often the only source of information to infer the strength of selection on that character. Time series of ancient DNA, in particular, are becoming available thanks to modern advances in preparation and sequencing methods \cite{Schraiber2015a, Malaspinas2016a}. These past records deliver the fluctuating frequency of an allele over time. The nature of these fluctuations is characterized by the combined effect of various mechanisms, the simplest of which are natural selection and genetic drift, on which we will focus our attention here. While natural selection drives the frequency towards fixation or stabilization, genetic drift caused by a small effective population size works towards elimination of genetic diversity and, thus, towards fixation of one of the characters or alleles \cite{Gillespie2010a}. As such, if the population size is known, genetic drift is a noisy effect that changes the frequency of the alleles and masks the effect of selection.

Natural selection can be theoretically described with relatively simple population genetics models, such as the {\red Moran and the Wright-Fisher} models \cite{Gillespie2010a, Ewens2012a}. At the basis of these models, the effect of natural selection is often crystallized in one single parameter per locus, called the selection coefficient. One of the tasks ahead of the analysis of DNA time-series is thus the extrapolation of the underlying selection coefficient. Indeed, the selective advantage of a certain character is quite impossible to determine from first principles, e.g.\ from an evaluation of metabolic costs and benefits, with the exception perhaps of a few experimentally controlled cases in bacterial populations. But even in bacteria, the advantage of a certain gene compared to another is determined indirectly, mostly by competition experiments and growth rate measurements\cite{Woods2011a}. 

Various methods, mostly based on maximum likelihood techniques have been developed to duly take both genetic drift and sampling errors into account \cite{Bollback2008a, Malaspinas2012a, Mathieson2013a, Feder2014a}. Several limiting cases have considered the task of determining the selection coefficient in the absence of genetic drift, \ie with large populations, thus taking in fact a deterministic approach\cite{Woods2011a, Illingworth2011a,  Illingworth2012a, Illingworth2014a}. The limiting {\red cases that we consider here are both} an haploid character with two competing alleles  {\red and a one-locus two-allele model with selection and codominance. We consider a finite population with perfect sampling. These conditions allow an analytic and precise treatment of the effect of genetic drift.} 

Taking apart those cases where the population size is too big for genetic drift to play any role, in the general case it is possible that the less advantageous character or allele is present at a larger frequency than a competing but more advantageous character. Nevertheless, we may inquire if and when a given time-series of the frequencies is informative of the relative selection strength of the two competing characters. {\red Simple models of population genetics}, albeit sometimes not completely realistic, provide a clear platform to derive analytical results easy to interpret and generalize. The aim of this work is to introduce a new likelihood function that works for any strength of the selection coefficient and for any value of the frequency, \ie also for frequencies close to the fixation boundary.
Accordingly, in order to understand the potentiality and the limits of such analysis we will {\red first work with} the Moran model of population genetics, which is by far one of the most intensively studied and successful metaphor of evolution under selection and drift \cite{Moran1958a, Lieberman2005a, Ewens2012a}. {\red We will then study the same problem with the one-locus two-allele Wright-Fisher model, which is definitively a more complex and more realistic metaphor of natural selection and drift\cite{Bollback2008a}}.

As we shall see, extracting the selection coefficient even in such a simple set-up is tricky. If one uses the wrong likelihood, apparently meaningful, self-consistent but otherwise incorrect conclusions are produced. The key point will be to understand that single time-series of processes that are {\em per se} non-stationary need to be treated as stochastic processes conditioned both in the initial and the final condition.

\section*{Results}

{\red The models that we consider have two types of alleles, $A$ and $B$. In the Moran model we will have haploid individuals carrying the alleles of type $A$ and $B$. In the Wright-Fisher model we will have diploid individuals carrying a pair of alleles of types $A$ and $B$ in one autosomal locus. Although these two models differ in structure and complexity, it is still possible to provide a common description of the underlying process of selection and drift. We start by considering a population of $N$ alleles. To allow for a common treatment of both models we will assume that $N$ is an even number. At any point in time, $N_A$ and $N_B$ are the number of alleles of type $A$ and $B$ in the population, respectively, and at each time point $N_A+N_B=N$ holds. We will say that $N_A$ and $N_B$ are the frequencies of alleles $A$ and $B$, respectively. Throughout the whole manuscript, we assume that these frequencies can be measured exactly (no sampling errors).

We will follow the fate of the number of alleles of type $B$ whose dynamics will be described as a Markov chain in discrete time with two absorbing states in $N_B=0$ and $N_B=N$. These two absorbing states correspond to the fixation of allele $A$ and $B$, respectively.
\begin{figure}[ht]
\captionsetup{labelformat=empty}
\begin{center}
\includegraphics[scale=0.5]{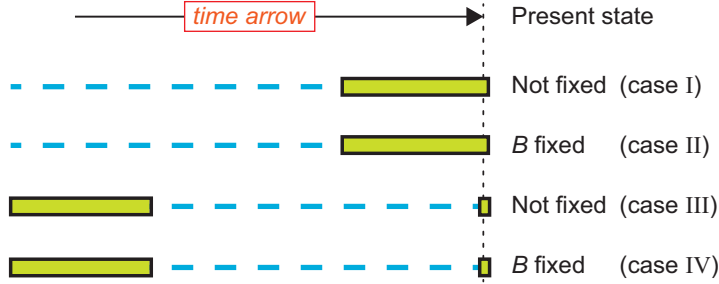}
\end{center}
\caption{{\bf Figure 1: {\red Four kinds of time series}}. {\red Schematic representation of the four cases considered here. The green bars represent available data for $T$ consecutive generations, whereas the blue dashed lines represent non available data. The time arrow goes from left to right with the present time called generation $T_F$. The data includes an initial condition at generation zero. We follow the trajectory of the allele $B$, whose frequency at present time is known in all cases. In cases $I$ and $II$, $T_F$ is just one generation after the measurement $T$, \ie $T_F=T+1$, so that the available data concern the recent history of the allele. In cases $III$ and $IV$ the measurement $T_F$ is made a long time after the measurement $T$. We can think of the cases $III$ and $IV$ as time-series where both the ancient history and the present frequency are known in detail but data in between are missing. Within the Wright-Fisher model we consider a variant of cases $II$ and $IV$, in which $B$ is very close to fixation but not yet fixed.} \label{fig1}}
\end{figure}

A single historical trajectory of $T+2$ measurements for the frequency $N_B$ can be used to estimate the selection coefficient. The trajectory has a initial condition $N_B(0)=i$, followed by $T$ intermediate measurements from strictly consecutive updates and one additional, final measurement at $T_F$. In what follows, while $N_B(0)$ is the same for all cases studied here, we consider various options for the timing $T_F$ of the final measurement and for the value of the frequency of the alleles of type $B$ at $T_F$, $N_B(T_F)$. We will also assume that the time is measured in generations, even if, strictly speaking, in the Moran model the generations are overlapping and in the Wright-Fisher model they are non-overlapping. We consider a total of four different limiting cases ({\bf Fig.\ \ref{fig1}})}. On the one hand, the first two cases are when $T_F$ is just one generation after the $T$th measurement, \ie $T_F=T+1$. Ideally, these first two cases correspond to time series of consecutive generations finishing at present time. Case $I$ is defined when $N_B(T_F)$ is at an intermediate frequency, \ie $N_B(T_F)\neq 0,N$. {\red Case $II$ is when $N_B(T_F)=N$, namely when the allele of type $B$ has reached fixation before or at present time.} On the other hand, the second two cases are when generation $T_F$ is long after generation $T$, \ie $T_F\gg T$. Ideally, this corresponds to trajectories where the initial time $t=0$ of the temporal observation is far back in the past so that also after $T$ generations the time-series {\red of available data} is still far back in the past. Here, generation $T_F$ is at present time and $N_B(T_F)$ is known but {\red the values of $N_B$ at times between generations $T$ and $T_F$ are missing. We then distinguish between case $III$, when the present frequency $N_B(T_F)$ is at any intermediate frequency, \ie $N_B(T_F)\neq 0,N$, {\red and case $IV$ where the present frequency is at fixation for the allele of type $B$, \ie  $N_B(T_F)=N$.} Obviously, cases $III$ and $IV$ reduce to cases $I$ and/or $II$ when the frequency at present time is ignored. As will become clear later, these cases are definitively different depending on the assumption  that one makes for the present state. One can also recognize that case $III$ is the most studied one in the literature so far\cite{Bollback2008a, Malaspinas2016a, Zhao2016a}. Since in cases $II$ and $IV$ fixation can occur at any generation including generations $t<T$, with the Wright-Fisher model we have also considered a variant of these two cases in which $N_B(T_F)=N-1$, \ie very close to fixation but not yet fixed. These variants do not present substantial differences in the results and are further discussed below.

For each one of these four cases we generate 100 independent time-series while keeping the selection coefficient fixed to $S=2$, whose meaning is explained below for each of the two models separately.  We generate such trajectories via stochastic simulations and then analyze them with the likelihood developed below to prove if we are able to reliably extract the selection coefficient. Within each of the four cases $I$ to $IV$, all trajectories share the same initial and final conditions $N_B(0)$ and $N_B(T_F)$, respectively, but are otherwise completely independent. 

Each trajectory is fully described by the index functional $\delta_{ij} (t)\in \{0,1\}$ such that
\beq
\delta_{ij}(t)\, =\, \left\{
\begin{array}{lcl}
1 & \mbox{ if } & N_B(t)=i\, \,  \mbox{ and }\, \, N_B(t+1)=j \\
& & \\
0 & & \mbox{otherwise}\, ,
\end{array}
\right.
\eeq
namely $\delta_{ij}(t)=1$ when a transition from frequency $i$ to frequency $j$ of the number of $B$ alleles occurs at time step $t$. The index $t$ runs over the measurements, $t=0,1,2,\ldots, T$. Thanks to this functional, the selection coefficient can be estimated by means of the conditioned likelihood
\beq
\label{TrueL}
L_c(s)\, =\, \prod_{t=0}^T \, \left[\prod_{i,j=0}^N \left( P_{ij\mid k}(s, T_F-t)^{\delta_{ij}(t)}\right)\right]\, ,
\eeq
where lowercase $s$ refers to the estimated value of $S$ and the conditional transition probability is defined as
\beq
P_{ij\mid k}(s, T_F-t)\, =\, \Pr\left\{ N_B(t+1)=j\mid N_B(t)=i,\,\, N_B(T_F)=k\right\}\, .
\eeq
The selection coefficient $s$ enters into this definition through the explicit form of the model as will be discussed in detail below and in the {\em Methods} section.

The application of the conditioning at the final frequency $N_B$ at the end of the trajectory allows to explicitly write the relationship between the conditioned and the non-conditioned transition probabilities by exploiting the Markov property of the chains, as\cite{Valleriani2015a} 
\beq
\label{pq}
P_{ij\mid k}(s, T_F-t)\, =\, \frac{\Pr\left\{N_B(T_F)=k\mid N_B(t+1)=j\right\}}{\Pr\left\{N_B(T_F)=k\mid N_B(t)=i\right\}}\Pr\left\{ N_B(t+1)=j\mid N_B(t)=i\right\}\, ,
 \eeq
 which in a shorthand we write as
 \beq
 \label{pqphi}
 P_{ij\mid k}(s, T_F-t)\, =\, \phi_{ij\mid k}(s, T_F-t) \cdot P_{ij}(s)\, ,
 \eeq
 where $P_{ij}(s)$ are the non-conditioned transition probabilities as defined by the model. }The functions $\phi_{ij\mid k}$ are instead complex functionals, determined by the Doob's $h$-transform, that depend on $s$, $i$, $j$ and $T_F-t$ ({\em Methods}).  

If we could ideally access a large number of trajectories collected under the same initial condition but free to {\red cover the available fitness landscape, only the initial condition would matter and the condition in the final state is no longer necessary.} This case is what one encounters in experimental evolution. The estimation of the selection coefficient in those cases should be made by means of the unconditioned likelihood\cite{Anderson1957a} {\red  
\beq
\label{like}
L(s)\, =\, \prod_{i,j=0}^{N} P_{ij}(s)^{n(\{i,j\})} \, ,
\eeq
where $n (\{i,j\})$ is the number of transitions between each pair of frequencies $i$ and $j$ in the ensemble of trajectories. The number of transitions can be computed through $n (\{i,j\})=\sum_t \delta_{ij}(t)$.} The likelihood function $L(s)$, and variations thereof that take sampling errors into account, is the most commonly used function to estimate the selection coefficient\cite{Malaspinas2016a}. {\red For a correct interpretation of the results presented below it is relevant to note that $L_c(s)$ and $L(s)$ are related through
\beq
L_c(s)\, =\, \Phi(s)\, L(s)\, ,
\eeq
where $\Phi(s)$ is a complex functional depending on the $\phi_{ij\mid k}$ and on the specific trajectory described by $\delta_{ij}(t)$. 

In the following we present a comparison of the estimated value of the selection coefficient from the likelihood $L_c(s)$ and from the likelihood $L(s)$, for the Moran and the Wright-Fisher models. Applied to each single trajectory both likelihoods allow to derive the most likely value of $s$. The variation of the maximum likelihood estimates across the set of 100 time-series for each of the four cases introduced earlier ({\bf Fig.\ \ref{fig1}}) gives a distribution from which the average and the 95\% confidence interval can be estimated.  } 

{\red
\subsection*{The Moran model}
\begin{figure}[ht]
\captionsetup{labelformat=empty}
\begin{center}
 \includegraphics[scale=0.5]{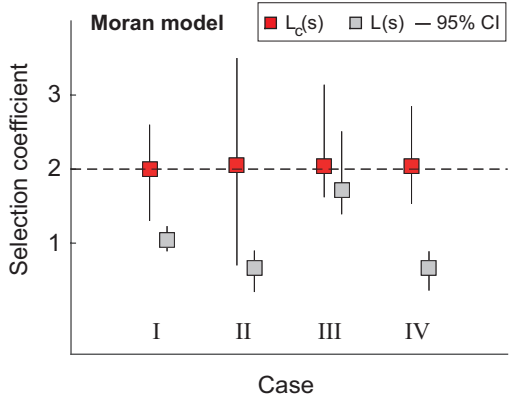}
\end{center}
\vspace{-2cm}
\caption{{\bf Figure 2: {\red Selection coefficient for the Moran model.}} {\red For each of the four cases, we have generated 100 independent trajectories with $S=2$ (dashed horizontal line). For each such trajectories we have constructed the likelihoods $L_c(s)$ and $L(s)$ and found the two values of $s$ that maximize each of them separately. From the distribution of these two sets of maximizing $s$ we obtain the mode and the 95\% confidence interval (CI) shown here. The conditioned likelihood $L_c(s)$ always provides a good estimate of the true selection coefficient (red squares). The unconstrained likelihood $L(s)$ delivers a poor estimate of the selection coefficient (grey squares) except for case $III$ due to the slow dynamics of the Moran model. For each trajectory: $T=500$, $N_B(0)=27$ and $N=54$. In cases $I$ and $III$ we have set $N_B(T_F)=40$. In cases $II$ and $IV$ we have set $N_B(T_F)=N$. }}
\label{fig2}
\end{figure}
 We consider the simplest version of the Moran model \cite{Moran1958a, Ewens2012a, Nowak2006a}, which consists of a population of $N$ individuals split into $N_A$ individuals carrying the character $A$ and $N_B$ individuals carrying the character $B$. Except for the characters $A$ and $B$, the individuals are identical. Individuals of type $A$ have fitness $W_A$ and individuals of type $B$ have fitness $W_B$. The selection coefficient is $S=W_A/W_B$. In the Moran model the generations are overlapping and the dynamics runs as follows. At each time point $t$, one of the existing individuals reproduces with a probability proportional to its fitness. The resulting offspring is identical to the parent individual and replaces one of the existing individuals chosen at random with uniform probability. At each time step, thus, the number $N_B$ of $B$ individuals can increase or decrease by one, or stay the same with probabilities that depend on $N_B$, $N$ and $S$ ({\em Methods}). The Moran model is thus a random walk on a line for the number $N_B$, with two absorbing states in $0$ and $N$ corresponding to the fixation of the character $A$ and $B$, respectively.
 
For this model, the 100 trajectories of type $B$ frequencies for each of the four types ({\bf Fig.\ \ref{fig1}}) have a duration of $T=500$ generations. The trajectories have been generated by standard methods for conditioned processes\cite{Valleriani2015a, Zhao2014a} and then both the conditioned likelihood $L_c(s)$ and the non-conditioned likelihoods $L(s)$ have been numerically derived as described. Surprisingly, only for the case $III$, \ie time-series far back in the past with the character not yet fixed at present time, also $L(s)$ delivers a selection coefficient close to the true one  ({\bf Fig.\ \ref{fig2}}, grey boxes). In all other cases $I$, $II$, and $IV$, $L(s)$ delivers selection coefficients that are quite different from the true one. 

\subsection*{The Wright-Fisher model for one-locus two alleles}
\begin{figure}[ht]
\captionsetup{labelformat=empty}
\begin{center}
\includegraphics[scale=0.5]{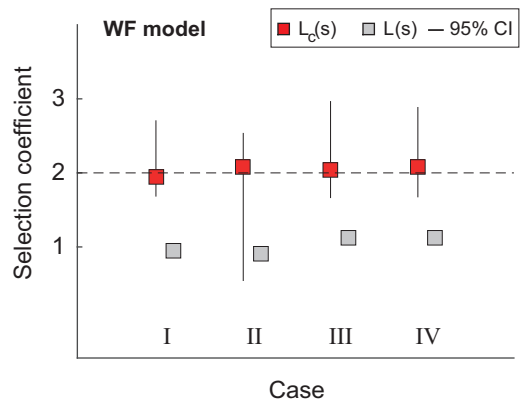}
\vspace{-2cm}
\end{center}
\caption{{\bf Figure 3: {\red Selection coefficient for the Wright-Fisher model.}} {\red For each of the four cases, we have generated 100 independent trajectories with $S=2$ (dashed horizontal line). For each such trajectories we have constructed the likelihoods $L_c(s)$ and $L(s)$ and found the two values of $s$ that maximize each of them separately. From the distribution of these two sets of maximizing $s$ we obtain the mode and the 95\% confidence interval (CI) shown here. Here, the rapid dynamics of the Wright-Fisher model leads to very short trajectories in cases $II$ and $IV$ that leads to poor statistics. For this reason, in cases $II$ and $IV$ we have set $N_B(T_F)=N-1$ (very close to fixation) instead of $N$. The conditioned likelihood $L_c(s)$ always provides a good estimate of the true selection coefficient (red squares). The unconstrained likelihood $L(s)$ delivers a poor estimate of the selection coefficient (grey squares) with a CI smaller than the box size. For each trajectory: $T=100$, $N_B(0)=27$ and $N=54$. In cases $I$ and $III$ we have set $N_B(T_F)=40$. }
\label{fig3}}
\end{figure}
In the Wright-Fisher model we consider an autosomal locus of a diploid organism with two alleles $A$ and $B$. Reproduction occurs with perfect mixing but population size is fixed to a total of $N$ alleles (corresponding to $N/2$ individuals). The three possible genotypes have fitness $W_{AA}$, $W_{BB}$ and $W_{AB}$. The selection coefficient is $S=W_{AA}/W_{BB}$ with codominance implying $W_{AB}/W_{BB} = (1+S)/2$. With these choices, in the absence of genetic drift the evolutionary trajectory would deterministically lead to the fixation of the allele $A$. For finite populations instead, the zygotes of the next generation are sampled from the gametes from the previous generation, in which the frequency of the alleles $A$ and $B$ have been determined through the evolutionary dynamics. The number $N_B$ of alleles of type $B$ in a finite adult population thus changes randomly from one generation to the next as a result of selection and drift ({\em Methods}). Also here the number of alleles $N_B$ is described as a Markov chain with two absorbing states in $0$ and $N$, corresponding to the fixation of allele $A$ and $B$, respectively.

This model is numerically more challenging than the Moran model. In particular, the time scale to fixation is shorter than for the Moran model because here the generations are non-overlapping. Here, thus, each trajectory has a duration of $T=100$ generations. As for the Moran model, we have generated 100 independent time-series for each of the four types ({\bf Fig. \ref{fig1}}). Using the transition probabilities of this model and the $\delta_{ij}(t)$, we have numerically derived the conditional likelihood $L_c(s)$ and the unconditioned likelihood $L(s)$. The results are qualitatively similar to the ones for the Moran model ({\bf Fig.\ \ref{fig3}}). For type $III$ trajectories, however, the two likelihoods perform differently, with $L_c(s)$ providing a good estimate of the selection coefficient and $L(s)$ a poor estimate. As discussed below, this has to do with the very rapid time scales of the Wright-Fisher model. If $T$ is set to 10 generations instead of 100, the estimate from $L(s)$ becomes closer to the true value. Due to its rapid time scales, it was also convenient to set $N_B(T_F)=N-1$, \ie very close to fixation, in order to have relatively long trajectories. }

\section*{Methods}
{\red In both models considered here, the process $N_B$ is a Markov chain in discrete time in a finite state space $\{0,1,\ldots , N\}$. These Markov chains are characterized by the one step transition probability matrix $\mathbbm{P}$ whose elements $P_{ij}$ are independent of time and are defined as 
\beq
\label{nakedP}
P_{ij}\, =\, \Pr\{N_B(t+1)=j\mid N_B(t)=i\}\, .
\eeq
The factors $\phi_{ij\mid k}$ that enter into the definition of the conditioned transition probabilities can be explicitly written by exploiting the definition of conditional probabilities and the Markov property of the process\cite{Valleriani2015a} as
\beq
\label{phis}
\begin{array}{lcl}
\bds \phi_{ij\mid k} (s, T_F-t)\, \eds & =\,&\bds \frac{\Pr\{N_B(T_F)=k\mid N_B(t+1)=j \}}{\Pr\{N_B(T_F)=k\mid N_B(t)=i\}}\,=\, \left(\mathbbm{P}^{T_F-t-1}\right)_{j k}/\left(\mathbbm{P}^{T_F-t}\right)_{ik}\eds
\end{array}
\eeq
which are non-negative functions dependent explicitly on $i$, $j$, $k$ and $T_F - t$ for $t=0,1,2\ldots, T$. When $T_F=T+1$, as in the cases $I$ and $II$ ({\bf Fig.\ \ref{fig2}}), the factors $\phi_{ij\mid k}$ depend explicitly on time and change in such a way to realize the condition $N_B(T_F)$. Nevertheless, the knowledge of the transition probabilities $P_{ij}$ defined in Eq.\ (\ref{nakedP}) allow to compute all likelihoods through Eq.\ (\ref{phis}) for any choice of the parameters. When $T_F\gg T$, as in the cases $III$ and $IV$ ({\bf Fig.\ \ref{fig1}}), the factors $\phi_{ij\mid k}$ do not depend on time\cite{Huillet2010a} and can be computed as the mathematical limit $T_F\to\infty$ by exploiting the spectral properties of the transition probability matrix $\mathbbm{P}$. When $k$ is a transient state, \ie $k\neq 0,N$, then 
\beq
\phi_{ij\mid k} (s)\, =\, \lambda_0^{-1}\,\frac{w_0(j)}{w_0(i)}\, ,
\eeq
where $\lambda_0$ is the largest non-trivial eigenvalue of $\mathbbm{P}$ and $w_0(i)$ is the $i$-th component of the corresponding right eigenvector. When $k$ represents fixation, \ie $k=0$ or $N$, then\cite{Valleriani2015a}
\beq
\phi_{ij\mid k} (s)\, =\, \frac{u_{jk}}{u_{ik}}\, ,
\eeq
where $u_{ik}$ is the probability of absorption in $k$ for a process started in $i$. Since deciding when $T_F$ is sufficiently large to allow using these last limit cases may depend on the system\cite{Zhao2016a}, the definition given in Eq.\ (\ref{phis}) was used to the limits of numerical precision for large powers.

For the Moran model, at each generation, each individual of type $A$ produces a number of offspring equal to $W_A$ and each individual of type $B$ produces a number of offspring equal to $W_B$. At each generation, just one among the entire pool of $N_A W_A+N_B W_B$ offspring is chosen at random. This new individual, then, replaces one randomly chosen individual in the parents' population. With this dynamics, the population size remains constant but the frequencies $N_A$ and $N_B$ change with time. Eventually, all individuals will be either of type $A$ or of type $B$. }

We follow the fate of character $B$. At each generation and before fixation occurs, the frequency $N_B$ can increase by one, decrease by one or stay the same. Based on the dynamics described above, the probabilities associated to the changes of $N_B$ are given by
\beq
\label{PQ}
\begin{array}{lcl}
\bds
P_i(S)\equiv \Pr\{N_B(t+1) = i+1\mid N_B(t)=i\}\eds & =
  &  \bds \frac{i(N-i)}{N(N-i)S+ iN} \eds \\
& & \\
\bds Q_i(S)\equiv \Pr\{N_B(t+1)=i-1\mid N_B(t)=i\} \eds  & = 
&  \bds \frac{i(N-i) S}{N(N-j)S + iN} \eds\\
& & \\
\bds R_i(S)\equiv \Pr\{N_B(t+1)=i\mid N_B(t)=i\}\eds & = & \bds 1 - P_i (S) - Q_i (S)\, ,\eds  
\end{array}
\eeq
where the selection coefficient $S=W_A/W_B$ is non-negative and the transition probabilities are independent of time.  When $0\le S <1$ the individuals of type $B$ have a selective advantage with respect to individuals of type $A$ (\ie $P_j>Q_j$) and {\em vice versa} when $S>1$. The borderline case $S=1$ corresponds to neutral evolution. The probabilities $P_i$, $Q_i$ and $R_i$ form the elements of the transition probability matrix 
\beq
\left(\mathbbm{P}\right)_{ij}\, =\, \left\{ \begin{array}{ccl}
Q_i(S) & \mbox{ if } & j=i-1,\, i\neq 0,N \\
R_i(S) & \mbox{ if } & j=i \\
P_i(S) & \mbox{ if } & j=i+1,\, i\neq 0,N\\
0 & & \mbox{otherwise}
\end{array}\right.
\eeq
The fixation probabilities as function of the initial frequencies and of the selection coefficient can be computed as absorption probabilities from this matrix\cite{Ewens2012a, Nowak2006a,Valleriani2015a}. 

{\red For the Wright-Fisher model, let $N_B(t) = i$ be the number of alleles of type $B$ in the adult population at generation $t$. Then, according to the evolutionary dynamics the frequency of the allele $B$ in the successive {\em gamete} population is\cite{Gillespie2010a}
\beq
\pi_B(i) \, =\, \frac{W_{BB}\,p_B(i)+W_{AB}\,p_A(i)}{W_O}\, p_B(i)\, ,
\eeq
where $p_B(i)=i/N$, $p_A(i)=1-p_B(i)$ and $W_O$ is the average fitness of the adult population, defined as
\beq
W_O \, =\, W_{AA}\, p_A^2(i)+2W_{AB}\, p_A(i)p_B(i)+W_{BB}\, p_B^2(i)\, .
\eeq
The frequency of the allele of type $B$ in the new adult population is obtained through random sampling and leads to the transition probabilities
\beq
P_{ij} \, =\, \left(\begin{array}{c} N \\ j \end{array}\right) \pi_B^j(i)(1-\pi_B(i))^{N-j}\, ,
\eeq
according to the binomial sampling. }

\section*{Discussion}
{\red At a first sight, it may seem odd that the correct likelihood should depend on the conditional transition probabilities $P_{ij\mid k}(s)$. In fact, $L_c(s)$ is computed on one single trajectory of a stochastic process governed by selection and genetic drift. The key point is that single trajectories of a stochastic process should be considered as representative of a bundle of trajectories starting {\em and} ending at fixed conditions. Functionals of single trajectories are thus conditioned not only in the initial condition but also in their final condition. When only one realization of ancient DNA variations is available a special form of conditioning in the future has to be included in order to correctly estimate the selection parameters. Such processes were already studied by Schr{\"o}dinger\cite{Schroedinger1931a} who recognized the emergence of possible contradictory claims from the observation of diffusion trajectories conditioned in their initial and final positions. In mathematics, this kind of conditioning has been studied in the context of {\em Brownian bridges}, namely processes conditioned both at their initial and final value, a precise description of which requires the introduction of the Doob's {\em h}-transform. More recently, the Doob's {\em h}-transform has become an essential theoretical tool to study the statistics of rare events\cite{Chetrite2014a} and to understand circular arguments in statistical analysis\cite{Valleriani2015a,Rusconi2016a}.
It was also shown that this transform emerges necessarily when trajectories are selected on the basis of their outcome\cite{Valleriani2015a}. }

{\red The likelihood $L(s)$ defined in Eq.\ (\ref{like}), based on the transition probabilities $P_{ij}$ given in Eq.\ (\ref{PQ}) is not the one that should be used to extract a parameter like the selection coefficient from one given trajectory. } Indeed, $L(s)$ fails in almost all cases to provide a realistic interval of confidence. The reason for the failure of this method is born in the fact that a given realization does not reveal if it is an unlikely event of a process that would otherwise typically behave differently. As a matter of fact, the process behind a given realization is rather more representative of a process conditioned (in probabilistic terms) to end at the frequency observed at its final observation. If one knows, from first principles, what is the microscopic (molecular) mechanism driving the process under scrutiny then one can follow the procedure explained in this work, derive the {\red conditional  probabilities $P_{ij\mid k}$} and write the likelihood $L_c(s)$ in terms of these conditioned quantities. This quite obviously provides a good estimate of the selection coefficient. A crucial requirement for the success of this enterprise is the knowledge of the correct model to use. 

The use of the {\red unconditioned} likelihood $L(s)$ would still give an answer, \ie a value of $s$ that is apparently consistent with the data. Indeed, case $I$, which describes a process conditioned on ending at an intermediate state $k\neq 0,\, N$ would lead to support the idea of neutral evolution or balancing selection {\red and in fact, $L(s)$ yields a value of $s$ close to unity}. In case $II$, when $k=N$ instead, the individuals of type $B$ would get fixed in the population and the analysis of such a trajectory by means of {\red unconditioned} likelihood $L(s)$ would lead to support the idea of a selective advantage in favor of type $B$ even if type $A$ individuals have a selective advantage by construction. Moreover, the time dependence of the transition probabilities, due simply to the effect of conditioning as seen in Eq.\ (\ref{phis}), would deceptively support the idea of changing environmental conditions. We see that these conclusions, albeit logical from the point of view of explaining the observations {\em a posteriori}, are determined by conditioning, \ie by the fact that $N_B(T_F)$ takes a particular value. Given our {\em a priori} knowledge of how we have generated the trajectories, conclusions taken through the analysis with the likelihood $L(s)$ would be therefore deprived of any foundation. But if we had no such {\em a priori} knowledge, there would be no way to confirm or reject the conclusions based on $L(s)$. {\red Case $III$, with data coming from far back in the past and no fixation, presents some peculiarity. For the Moran model $L(s)$ gives a relatively good estimate of the true selection coefficient whereas for the Wright-Fisher model it does not. The reason relies on the different time scales associated to absorption in each of these models. One step in the Wright-Fisher model corresponds to at least $N$ steps in the Moran model. Thus, when the duration $T$ of the time-series is very long and no absorption takes place at the end or close to the end of the time-series, the analysis performed with $L(s)$ leads to a value of $s$ close to unity, compatible with the apparent neutrality of the evolutionary trajectory. When $T$ is short, instead, also for the Wright-Fisher model $L(s)$ delivers a value of $s$ closer to the true value (a test done with $T=10$ confirmed this assertion). Therefore, the effect of conditioning in the future combined with the typical time scale of the process and the length of the measurement $T$ is non trivial\cite{Zhao2016a}. Finally, in case $IV$} conditioning can be very strong because the process can enter fixation at any time before the present, including times during the observed time-series. From the point of view of the likelihood $L(s)$, case $IV$ would give type $B$ individuals a selective advantage where $L_c(s)$ instead correctly predicts that $A$ was in advantage. {\red Furthermore, in the light of the relationship between $L_c(s)$ and $L(s)$, it emerges especially in trajectories belonging to case $II$ that $L_c(s)$ is bimodal, with a local maximum governed by $L(s)$ and a second local maximum at larger values of $s$ governed by $\Phi(s)$. This explains the larger confidence interval for this case in both models. This suggests that the ratio of the likelihoods $R(s)=L_c(s)/L(s)$ rather than $L_c(s)$ alone could be considered an even better functional to estimate the true value of the selection coefficient. }

It had already been observed in the context of {\red other models} of population genetics that the generation of faithful trajectories of allele frequencies under the condition that fixation has occurred requires the introduction of a fictitious selection coefficient\cite{Zhao2013a, Zhao2014a}. In the context of the Moran model instead, it was shown that under the condition that fixation has not occurred after long time, the transition probabilities require a correction factor\cite{Huillet2010a}. While both these cases are included and generalized in this manuscript, we should stress here instead that extrapolating the selection coefficients from single historic records without due consideration to the peculiar conditioning associated to single trajectories gives values of the selection coefficients that are often very different from the real values.

\section*{Conclusions}
An historic time-series is one trajectory whose contingency acts as a condition in the future and thus enters in the form of a bias in the elementary transition probabilities. The existence of such a bias when processes are conditioned in the future is often referred to as the Doob's {\em h}-transform. Extracting the selection coefficient from frequency time-series using the false likelihood function has a deceptive effect: the extracted parameters seem to be meaningful and would support models that completely agree with the data used to extract them. Especially when predictions about the future outcomes are not possible because of the experimental limitations, seeking for models solely from past macroscopic data generates a false self-consistency reminiscent of circularity in data analysis\cite{Kriegeskorte2009a, Brenner2010a, Valleriani2015a, Lewontin1991a}. When the correct model is known, it is possible to derive a likelihood function that takes the Doob's {\em h}-transform into account and to produce reliable estimates of the selection coefficient.

\begin{thebibliography}{10}
\expandafter\ifx\csname url\endcsname\relax
  \def\url#1{\texttt{#1}}\fi
\expandafter\ifx\csname urlprefix\endcsname\relax\def\urlprefix{URL }\fi
\providecommand{\bibinfo}[2]{#2}
\providecommand{\eprint}[2][]{\url{#2}}

\bibitem{Schraiber2015a}
\bibinfo{author}{Schraiber, J.~G.} \& \bibinfo{author}{Akey, J.~M.}
\newblock \bibinfo{title}{Methods and models for unravelling human evolutionary
  history}.
\newblock \emph{\bibinfo{journal}{Nature Reviews Genetics}}
  (\bibinfo{year}{2015}).

\bibitem{Malaspinas2016a}
\bibinfo{author}{Malaspinas, A.-S.}
\newblock \bibinfo{title}{{Methods to characterize selective sweeps using time
  serial samples: An ancient DNA perspective.}}
\newblock \emph{\bibinfo{journal}{Mol Ecol}} \textbf{\bibinfo{volume}{25}},
  \bibinfo{pages}{24--41} (\bibinfo{year}{2016}).

\bibitem{Gillespie2010a}
\bibinfo{author}{Gillespie, J.~H.}
\newblock \emph{\bibinfo{title}{Population {G}enetics: {A} concise guide}}
  (\bibinfo{publisher}{JHU Press}, \bibinfo{year}{2010}).

\bibitem{Ewens2012a}
\bibinfo{author}{Ewens, W.~J.}
\newblock \emph{\bibinfo{title}{Mathematical Population Genetics 1:
  {T}heoretical Introduction}}, vol.~\bibinfo{volume}{27}
  (\bibinfo{publisher}{Springer Science \& Business Media},
  \bibinfo{year}{2012}).

\bibitem{Woods2011a}
\bibinfo{author}{Woods, R.~J.} \emph{et~al.}
\newblock \bibinfo{title}{{Second-order selection for evolvability in a large
  Escherichia coli population}}.
\newblock \emph{\bibinfo{journal}{Science}} \textbf{\bibinfo{volume}{331}},
  \bibinfo{pages}{1433--1436} (\bibinfo{year}{2011}).

\bibitem{Bollback2008a}
\bibinfo{author}{Bollback, J.~P.}, \bibinfo{author}{York, T.~L.} \&
  \bibinfo{author}{Nielsen, R.}
\newblock \bibinfo{title}{Estimation of {$2N_e s$} from temporal allele
  frequency data}.
\newblock \emph{\bibinfo{journal}{Genetics}} \textbf{\bibinfo{volume}{179}},
  \bibinfo{pages}{497--502} (\bibinfo{year}{2008}).

\bibitem{Malaspinas2012a}
\bibinfo{author}{Malaspinas, A.-S.}, \bibinfo{author}{Malaspinas, O.},
  \bibinfo{author}{Evans, S.~N.} \& \bibinfo{author}{Slatkin, M.}
\newblock \bibinfo{title}{Estimating allele age and selection coefficient from
  time-serial data}.
\newblock \emph{\bibinfo{journal}{Genetics}} \textbf{\bibinfo{volume}{192}},
  \bibinfo{pages}{599--607} (\bibinfo{year}{2012}).

\bibitem{Mathieson2013a}
\bibinfo{author}{Mathieson, I.} \& \bibinfo{author}{McVean, G.}
\newblock \bibinfo{title}{Estimating selection coefficients in spatially
  structured populations from time series data of allele frequencies}.
\newblock \emph{\bibinfo{journal}{Genetics}} \textbf{\bibinfo{volume}{193}},
  \bibinfo{pages}{973--984} (\bibinfo{year}{2013}).

\bibitem{Feder2014a}
\bibinfo{author}{Feder, A.~F.}, \bibinfo{author}{Kryazhimskiy, S.} \&
  \bibinfo{author}{Plotkin, J.~B.}
\newblock \bibinfo{title}{Identifying signatures of selection in genetic time
  series}.
\newblock \emph{\bibinfo{journal}{Genetics}} \textbf{\bibinfo{volume}{196}},
  \bibinfo{pages}{509--522} (\bibinfo{year}{2014}).

\bibitem{Illingworth2011a}
\bibinfo{author}{Illingworth, C.~J.} \& \bibinfo{author}{Mustonen, V.}
\newblock \bibinfo{title}{Distinguishing driver and passenger mutations in an
  evolutionary history categorized by interference}.
\newblock \emph{\bibinfo{journal}{Genetics}} \textbf{\bibinfo{volume}{189}},
  \bibinfo{pages}{989--1000} (\bibinfo{year}{2011}).

\bibitem{Illingworth2012a}
\bibinfo{author}{Illingworth, C.~J.}, \bibinfo{author}{Parts, L.},
  \bibinfo{author}{Schiffels, S.}, \bibinfo{author}{Liti, G.} \&
  \bibinfo{author}{Mustonen, V.}
\newblock \bibinfo{title}{Quantifying selection acting on a complex trait using
  allele frequency time series data}.
\newblock \emph{\bibinfo{journal}{Molecular Biology and Evolution}}
  \textbf{\bibinfo{volume}{29}}, \bibinfo{pages}{1187--1197}
  (\bibinfo{year}{2012}).

\bibitem{Illingworth2014a}
\bibinfo{author}{Illingworth, C.~J.}, \bibinfo{author}{Fischer, A.} \&
  \bibinfo{author}{Mustonen, V.}
\newblock \bibinfo{title}{Identifying selection in the within-host evolution of
  influenza using viral sequence data}.
\newblock \emph{\bibinfo{journal}{PLoS Comput Biol}}
  \textbf{\bibinfo{volume}{10}}, \bibinfo{pages}{e1003755}
  (\bibinfo{year}{2014}).

\bibitem{Moran1958a}
\bibinfo{author}{Moran, P. A.~P.}
\newblock \bibinfo{title}{Random processes in genetics}.
\newblock In \emph{\bibinfo{booktitle}{Mathematical Proceedings of the
  Cambridge Philosophical Society}}, vol.~\bibinfo{volume}{54},
  \bibinfo{pages}{60--71} (\bibinfo{organization}{Cambridge Univ Press},
  \bibinfo{year}{1958}).

\bibitem{Lieberman2005a}
\bibinfo{author}{Lieberman, E.}, \bibinfo{author}{Hauert, C.} \&
  \bibinfo{author}{Nowak, M.~A.}
\newblock \bibinfo{title}{Evolutionary dynamics on graphs}.
\newblock \emph{\bibinfo{journal}{Nature}} \textbf{\bibinfo{volume}{433}},
  \bibinfo{pages}{312--316} (\bibinfo{year}{2005}).

\bibitem{Zhao2016a}
\bibinfo{author}{Zhao, L.}, \bibinfo{author}{Lascoux, M.} \&
  \bibinfo{author}{Waxman, D.}
\newblock \bibinfo{title}{{An informational transition in conditioned Markov
  chains: {A}pplied to genetics and evolution}}.
\newblock \emph{\bibinfo{journal}{{Journal of Theoretical Biology}}}
  \textbf{\bibinfo{volume}{402}}, \bibinfo{pages}{158--170}
  (\bibinfo{year}{2016}).

\bibitem{Valleriani2015a}
\bibinfo{author}{Valleriani, A.}
\newblock \bibinfo{title}{Circular analysis in complex stochastic systems}.
\newblock \emph{\bibinfo{journal}{Scientific Reports}}
  \textbf{\bibinfo{volume}{5}}, \bibinfo{pages}{17986} (\bibinfo{year}{2015}).

\bibitem{Anderson1957a}
\bibinfo{author}{Anderson, T.~W.} \& \bibinfo{author}{Goodman, L.~A.}
\newblock \bibinfo{title}{Statistical inference about {Markov} chains}.
\newblock \emph{\bibinfo{journal}{The Annals of Mathematical Statistics}}
  \bibinfo{pages}{89--110} (\bibinfo{year}{1957}).

\bibitem{Nowak2006a}
\bibinfo{author}{Nowak, M.~A.}
\newblock \emph{\bibinfo{title}{{Evolutionary Dynamics: Exploring the Equations
  of Life}}} (\bibinfo{publisher}{{Harvard University Press}},
  \bibinfo{year}{2006}).

\bibitem{Zhao2014a}
\bibinfo{author}{Zhao, L.}, \bibinfo{author}{Lascoux, M.} \&
  \bibinfo{author}{Waxman, D.}
\newblock \bibinfo{title}{{Exact simulation of conditioned Wright--Fisher
  models}}.
\newblock \emph{\bibinfo{journal}{Journal of Theoretical Biology}}
  \textbf{\bibinfo{volume}{363}}, \bibinfo{pages}{419--426}
  (\bibinfo{year}{2014}).

\bibitem{Huillet2010a}
\bibinfo{author}{Huillet, T.}
\newblock \bibinfo{title}{Siegmund duality with applications to the neutral
  {Moran} model conditioned on never being absorbed}.
\newblock \emph{\bibinfo{journal}{Journal of Physics A: Mathematical and
  Theoretical}} \textbf{\bibinfo{volume}{43}}, \bibinfo{pages}{375001}
  (\bibinfo{year}{2010}).

\bibitem{Schroedinger1931a}
\bibinfo{author}{Schr{\"o}dinger, E.}
\newblock \bibinfo{title}{{\"U}ber die {U}mkehrung der {N}aturgesetze}.
\newblock \emph{\bibinfo{journal}{Sitzungsber. Preuss. Akad. Wiss., Phys.-Math.
  Kl.}} \bibinfo{pages}{412--422} (\bibinfo{year}{1931}).

\bibitem{Chetrite2014a}
\bibinfo{author}{Chetrite, R.} \& \bibinfo{author}{Touchette, H.}
\newblock \bibinfo{title}{Nonequilibrium {M}arkov processes conditioned on
  large deviations}.
\newblock \emph{\bibinfo{journal}{Annales Henri Poincar{\'e}}}
  \bibinfo{pages}{1--53} (\bibinfo{year}{2014}).

\bibitem{Rusconi2016a}
\bibinfo{author}{Rusconi, M.} \& \bibinfo{author}{Valleriani, A.}
\newblock \bibinfo{title}{Predict or classify: The deceptive role of
  time-locking in brain signal classification}.
\newblock \emph{\bibinfo{journal}{Scientific Reports}}
  \textbf{\bibinfo{volume}{6}}, \bibinfo{pages}{28236} (\bibinfo{year}{2016}).

\bibitem{Zhao2013a}
\bibinfo{author}{Zhao, L.}, \bibinfo{author}{Lascoux, M.},
  \bibinfo{author}{Overall, A.~D.} \& \bibinfo{author}{Waxman, D.}
\newblock \bibinfo{title}{{The characteristic trajectory of a fixing allele: A
  consequence of fictitious selection that arises from conditioning}}.
\newblock \emph{\bibinfo{journal}{Genetics}} \textbf{\bibinfo{volume}{195}},
  \bibinfo{pages}{993--1006} (\bibinfo{year}{2013}).

\bibitem{Kriegeskorte2009a}
\bibinfo{author}{Kriegeskorte, N.}, \bibinfo{author}{Simmons, W.~K.},
  \bibinfo{author}{Bellgowan, P.~S.} \& \bibinfo{author}{Baker, C.~I.}
\newblock \bibinfo{title}{Circular analysis in systems neuroscience: {T}he
  dangers of double dipping}.
\newblock \emph{\bibinfo{journal}{Nature neuroscience}}
  \textbf{\bibinfo{volume}{12}}, \bibinfo{pages}{535--540}
  (\bibinfo{year}{2009}).

\bibitem{Brenner2010a}
\bibinfo{author}{Brenner, S.}
\newblock \bibinfo{title}{Sequences and consequences}.
\newblock \emph{\bibinfo{journal}{Philosophical Transactions of the Royal
  Society of London B: Biological Sciences}} \textbf{\bibinfo{volume}{365}},
  \bibinfo{pages}{207--212} (\bibinfo{year}{2010}).

\bibitem{Lewontin1991a}
\bibinfo{author}{Lewontin, R.~C.}
\newblock \bibinfo{title}{Facts and the factitious in natural sciences}.
\newblock \emph{\bibinfo{journal}{Critical {I}nquiry}}
  \textbf{\bibinfo{volume}{18}}, \bibinfo{pages}{140--153}
  (\bibinfo{year}{1991}).

\end{thebibliography}

\end{document}